\documentclass[ aps,%
floatfix,%
final,%
notitlepage,%
oneside,%
onecolumn,%
nobibnotes,%
nofootinbib,%
superscriptaddress,%
showpacs,%
]%
{revtex4}

\usepackage{epsfig}
\usepackage{amsfonts}
\usepackage{amsmath}
\usepackage{epsfig}
\usepackage{graphics}

\newcommand{\beq}{\begin{eqnarray}}\newcommand{\eeq}{\end{eqnarray}}
\newcommand{\beqa}{\begin{eqnarray*}}\newcommand{\eeqa}{\end{eqnarray*}}

\begin{document}

\title{ The study of leading twist light cone wave function of $\eta_c$ meson.}
\author{V.V. Braguta}
\email{braguta@mail.ru}

\author{A.K. Likhoded}
\email{Likhoded@ihep.ru}

\author{A.V. Luchinsky}
\email{Alexey.Luchinsky@ihep.ru}
\affiliation{Institute for High Energy Physics, Protvino, Russia}

\begin{abstract}
This paper is devoted to the study of leading twist light cone wave function of $\eta_c$ meson. 
The moments of this wave function have been calculated within three approaches: potential models, 
nonrelativistic QCD and QCD sum rules. Using the results obtained within these approaches 
the model for the light cone wave function of leading twist has been proposed. Being scale dependent
light cone wave function has very interesting properties at scales $\mu> m_c$: improvement of the 
accuracy of the model, appearance of relativistic tail and violation of nonrelativistic QCD velocity scaling rules. 
The last two properties are the properties of real leading twist light cone wave function 
of $\eta_c$ meson.
\end{abstract}


\maketitle

\newcommand{\ins}[1]{\underline{#1}}
\newcommand{\subs}[2]{\underline{#2}}
\vspace*{-1.cm}
\section{Introduction}

Commonly exclusive charmonium production at high energies is studied within 
nonrelativistic QCD (NRQCD) \cite{Bodwin:1994jh}. In the framework of this approach charmonium is considered as 
a bound state of quark-antiquark pair moving with small relative velocity $v \ll 1$. 
Due to the presence of small parameter $v$ the amplitude of charmonium production can be built 
as an expansion in relative velocity $v$. 

Thus in the framework NRQCD the amplitude of any process is a series in relative velocity $v$. Usually,
in the most of applications of NRQCD, the consideration is restricted by the leading order approximation 
in relative velocity. However, this approximation has two problems which make it unreliable. The first problem is connected with 
rather large value of relative velocity for charmonium $v^2 \sim 0.3,~v \sim 0.5$. For this value 
of $v^2$ one can expect large contribution from relativistic corrections in any process. So in any 
process resummation of relativistic corrections should be done 
or one should prove that resummation of all terms is not crucial. The second problem is connected with
QCD radiative corrections. The point is that due to the presence of large energy scale $Q$ there appear 
large logarithmic terms $(\alpha_s \log Q/m_c )^n, ~ Q \gg m_c $ which can be even more 
important than relativistic corrections at sufficiently large energy ( $Q \sim 10$ GeV). 
So this terms should also be resummed. In principle, it is possible to resum large logarithms
in the NRQCD factorization framework \cite{Petrelli:1999rh, Bodwin:2005hm}, however 
such resummation is done rarely. 

The illustration of all mentioned facts is the process of double charmonium production in $e^+ e^-$ annihilation at B-factories, 
where leading order NRQCD predictions \cite{Braaten:2002fi, Liu:2002wq, Liu:2004ga} are approximately by an order of magnitude less than experimental 
results \cite{Abe:2004ww,Aubert:2005tj}. The calculation of QCD radiative corrections \cite{Zhang:2005ch} diminished this
disagreement but did not remove it. Probably the agreement with the experiments can be 
achieved if, in addition to QCD radiative corrections, relativistic corrections will be resummed \cite{Bodwin:2006ke}.

In addition to NRQCD, hard exclusive processes can be studied in the framework of light cone expansion 
formalism \cite{Lepage:1980fj, Chernyak:1983ej} where both problems mentioned above can be solved. Within 
light cone expansion formalism the amplitude is built as an expansion 
over inverse powers of characteristic energy of the process. Usually this approach is successfully applied 
for excusive production of light mesons \cite{Chernyak:1983ej}. However recently the application of light cone 
expansion formalism to double charmonium production \cite{Bondar:2004sv, Braguta:2006nf, Braguta:2005kr, Ma:2004qf} allowed one
to achieve good agreement with the experiments.

In the framework of light cone formalism the amplitude of some meson production in any hard process  can be written as 
a convolution of the hard part of the process, which can be calculated using perturbative QCD,
and process independent light cone wave function (LCWF) of this meson that parameterizes nonperturbative effects.
From this one can conclude that charmonium LCWFs are key ingredient of any hard exclusive process with charmonium production. 
This paper is devoted to the study of leading twist LCWF of $\eta_c$ meson. 

The paper is organized as follows. In the next section all definitions needed in our calculation will be given. 
In Section III the moments of LCWF will be calculated in the framework of Buchmuller-Tye and Cornell potential models. 
Section VI is devoted to the calculation of the moments within NRQCD. QCD sum rules will be applied 
to the calculation of the moments in Section V. Using the results obtained in Sections III-V 
the model for LCWF will be built in Section VI. In Section VII this model will be compared 
with some other models proposed in literature. In the last section we summarize the results of our paper.

\section{ Definitions.}

The leading twist light cone wave function (LCWF) of $\eta_c$ meson can be defined as follows \cite{Chernyak:1983ej}
\beq
{\langle 0| {\bar Q} (z) \gamma_{\alpha} \gamma_5 [z,-z] Q(-z) | \eta_c \rangle}_{\mu}=i
f_{\eta_c} p_{\alpha} \int^1_{-1} d \xi \,e^{i(pz) \xi}
\phi ( \xi, \mu),
\label{lwf}
\eeq
where the following designations are used: $x_1, x_2$ are the parts of momentum of the whole meson carried by quark and antiquark 
correspondingly, $\xi = x_1 - x_2$, $p$ is a momentum of $\eta_c$ meson, $\mu$ is an energy scale. The factor $[z,-z]$, that makes matrix element (\ref{lwf})
gauge invariant, is defined as 
\beq
[z, -z] = P \exp[i g \int_{-z}^z d x^{\mu} A_{\mu} (x) ].
\eeq
The LCWF $\phi (\xi, \mu)$ is normalized as
\beq
\int_{-1}^1 d \xi \phi (\xi, \mu) =1. 
\label{norm}
\eeq
With this normalization condition the constant $f_{\eta_c}$ is defined as
\beq
\langle 0| {\bar Q} (0) \gamma_{\alpha} \gamma_5 Q(0) |\eta_c \rangle=i
f_{\eta_c} p_{\alpha}.
\label{def}
\eeq
LCWF $\phi (x, \mu)$ can be expanded \cite{Chernyak:1983ej} in Gegenbauer polynomials $C_n^{3/2}( \xi)$ as follows
\begin{eqnarray}
\phi (\xi, \mu) = \frac 3 4 (1 - \xi^2) \biggl [  1 + \sum_{n=2,4..} a_n(\mu) C_n^{3/2} ( \xi ) \biggr ].
\label{conf_exp}
\end{eqnarray}
At leading logarithmic accuracy the coefficients $a_n$ are renormalized multiplicatively 
\begin{eqnarray}
a_n(\mu) = \biggl ( \frac {\alpha_s( \mu)} {\alpha_s( \mu_0)} \biggr )^{\frac {\epsilon_n} {b_0}} a_n(\mu_0),
\label{ren}
\end{eqnarray}
where 
\beq
\epsilon_n=\frac 4 3 \biggl ( 1- \frac 2 {(n+1) (n+2)} + 4 \sum_{j=2}^{n+1} \frac 1 j \biggr ), ~~ b_0= 11 - \frac 2 3 n_{\rm fl}.
\label{an_dim}
\eeq
It should be noted here that conformal expansion (\ref{conf_exp}) is a solution of Bethe-Salpeter equation 
with one gluon exchange kernel \cite{Lepage:1980fj}.

From equations (\ref{conf_exp})-(\ref{an_dim}) it is not difficult to see that at infinitely large energy scale $\mu \to \infty$ 
LCWF $\phi (\xi, \mu)$ tends to the asymptotic form $\phi_{as} (\xi) = 3/4 ( 1- \xi^2 )$. But at energy 
scales accessible at current experiments the LCWF $\phi( \xi, \mu)$ is far from it's asymptotic form. The main 
goal of this paper is to calculate the LCWF $\phi(\xi, \mu)$ of $\eta_c$ meson.
One way to do this is to calculate the coefficients of conformal expansion (\ref{conf_exp}) $a_n$. 
Having this coefficients  at some energy scale $\mu_0$ one can find the LCWF $\phi( \xi, \mu)$ at any scale $\mu$. 
Alternative method to parameterize LCWF is to evaluate it's moments  at some scale
\beq
\langle \xi^n  \rangle_{\mu} = \int_{-1}^1 d \xi ~ \xi^n \phi (\xi, \mu).
\eeq
In our paper this method will be used. It is worth noting that since $\eta_c$ mesons have positive 
charge parity the LCWF $\phi( \xi, \mu)$ is $\xi$-even. Thus all odd moments 
$\langle \xi^{2 k+1} \rangle$ equal zero and one needs to calculate only even moments.

Below we will need the formula that connects moment $\langle \xi^n \rangle$ with 
the matrix element $\langle 0 |\bar Q \gamma_{\nu} \gamma_5 (i z^{\sigma} {\overset {\leftrightarrow} {D}}_{\sigma} )^n Q| P(p) \rangle$. 
To obtain it we expand both sides of equation (\ref{lwf}) 
\beq
\sum_n \frac {i^n} {n!} \langle 0 | \bar Q \gamma_{\nu} \gamma_5 (i z^{\sigma}  {\overset {\leftrightarrow} {D}}_{\sigma} )^n Q| \eta_c \rangle = i
f_{\eta_c} p_{\nu} \sum_n \frac {i^n} {n!} (zp)^n \langle \xi^n \rangle,
\eeq
and get
\beq
\langle 0 | \bar Q \gamma_{\nu} \gamma_5 (i z^{\sigma}  {\overset {\leftrightarrow} {D}}_{\sigma} )^n Q| \eta_c \rangle = i
f_{\eta_c} p_{\nu} (zp)^n \langle \xi^n \rangle.
\label{xin}
\eeq
Here
\beq
{\overset {\leftrightarrow} {D}}={\overset {\rightarrow} {D}}-{\overset {\leftarrow} {D}}, ~~~ {\overset {\rightarrow} {D}} = {\overset {\rightarrow} {\partial}}- 
i g B^a (\lambda^a /2).
\eeq

\section{ The moments in the framework of potential models. }

In potential models charmonium mesons are considered as a quark-antiquark system
bounded by some static potential. These models allow one to understand many properties 
of chamonium mesons. For instance, the spectrum of charmonium family can be very well 
reproduced in the framework of potential models \cite{Swanson:2006st}. Due to this 
success one can hope that potential models can be applied to the calculation of charmonium 
equal time wave functions. 

Having  equal time wave function of $\eta_c$ meson in momentum space $\psi( {\bf k} )$ one can apply Brodsky-Huang-Lepage (BHL) \cite{Brodsky:1981jv} procedure 
and get the LCWF of leading twist $\phi(\xi, \mu)$ using the following rule:
\beq
\phi (\xi, \mu) \sim \int^{{\bf k}_{\perp}^2< \mu^2} d^2 k_{\perp} \psi_c(x,{\bf k}_{\perp}),
\label{p1}
\eeq
where $\psi_c(x,{\bf k}_{\perp})$ can be  obtained from $\psi( {\bf k} )$ after the 
substitution \cite{Brodsky:1981jv}
\beq
{\bf k}_{\perp} \to {\bf k}_{\perp}, \quad k_z \to ( x_1 - x_2) \frac {M_0} 2, \quad M_0^2 = \frac {M_c^2 + {\bf k}_{\perp}^2 } {x_1 x_2}.
\label{sub}
\eeq
Here $M_c$ is a quark mass in potential model. 
In our paper equal time wave function $\psi( {\bf k})$ will be calculated in the framework of the potential models with 
Buchmuller-Tye \cite{Buchmuller:1980su} and Cornell potentials \cite{Eichten:1978tg}. The parameters of
Buchmuller-Tye potential model were taken from paper \cite{Buchmuller:1980su}. For Cornell potential $V(r) = -k/r+ r/a^2$
the calculation was carried our with the following set of parameters: $k=0.61,~ a=2.38 \mbox{~GeV}^{-1},~ M_c=1.84$ GeV \cite{Eichten:2002qv}. 

It is worth noting that in paper \cite{Bodwin:2006dm} the relations between the light cone wave functions 
and equal time wave functions of charmonium mesons in the rest frame were derived. The procedure proposed in paper \cite{Bodwin:2006dm}
is similar to BHL with the difference: in formula (\ref {p1}) one must make the substitution 
$d^2 k_{\perp} \to d^2 k_{\perp} \sqrt {{\bf k}^2+m_c^2}/(4 m_c x_1 x_2)$. But this substitution was derived at leading 
order approximation in relative velocity of quark-antiquark motion inside the charmonium. At this approximation 
${\bf k}^2 \sim O(v^2),~ 4 x_1 x_2 \sim 1+ O(v^2)$ and the substitution amounts to $d^2 k_{\perp} \to d^2 k_{\perp} (1 + O( v^2 ))$. 
Thus at leading order approximation applied in \cite{Bodwin:2006dm} these two approaches coincide. 

The potential models with Buchmuller-Tye and Cornell potentials treat charmonium as a nonrelativistic bound 
state, so these models can not be applied in the region where the motion is relativistic. From this one can conclude 
that it is not possible to apply expression (\ref{p1}) at sufficiently large scale $\mu \gg M_c$. At the same time conformal expansion (\ref{conf_exp})
can be applied at scales $\mu^2 \gg q^2 \sim M_c^2 v^2$ where $q$ and $v$ are characteristic momentum 
and velocity of $c \bar c$ system. Thus to find the values of moments $\langle \xi^n \rangle$ one should apply formula (\ref{p1}) 
at scale not too low and not too large. Our calculation will be done at 
scale $\mu = 1.5~ \mbox{GeV} \sim M_c$ where both conformal expansion ({\ref{conf_exp}}) and formula (\ref{p1}) can be applied. 
To estimate the scale dependence of our results in addition to scale $\mu=1.5$ GeV we have also done the calculation at scale $\mu=2$ GeV. 
The moments at scale $\mu=2.0$ GeV differ by few percents from the moments calculated at scale $\mu= 1.5$ GeV so the 
scale dependence is very weak.

The results of our calculation are presented in Table I. In second and third columns the moments calculated in the framework of 
Buchmuller-Tye and Cornell models are presented. It is seen that there is  good agreement between these two 
models. 

It should be noted here that the larger the power of the moment the larger contribution form 
the end point regions ($x \sim 0$ and $x \sim 1$) it gets. From formulas (\ref{sub}) one sees
that the motion of quark-antiquark pair in these regions is relativistic and cannot be considered 
reliably in the framework of potential models. Thus it is not possible to calculate higher moments 
within the potential models. Due to this fact we have restricted our calculation by few first moments.

\section{ The moments in the framework of NRQCD.  }

The moments of LCWF (\ref{lwf}) can be calculated in the framework of NRQCD. 
To do this let us consider matrix element $\langle 0 | \bar Q \gamma_{\nu} \gamma_5 (i {\overset {\leftrightarrow} {D}}_{\mu_1}) (i {\overset {\leftrightarrow} {D}}_{\mu_2}) Q |\eta_c \rangle$ 
in the meson's center mass frame. In this section we will work at leading order approximation in relative velocity of quark-antiquark 
pair $v$. According to the velocity scaling rules \cite{Bodwin:1994jh} gauge-covariant time derivative ${\overset {\leftrightarrow} {D}}_t$
is suppressed by one power of $v$ in comparison with gauge-covariant spatial derivative ${\overset {\leftrightarrow} {D}}_i$. Thus if we set $\mu_1$ or 
$\mu_2$ to zero than at leading order approximation this matrix element is zero. Moreover in the meson's center mass frame
only zero component of the current $\nu=0$ differs from zero. Regarding these two properties it causes no difficulties 
to built the matrix element:
\beq
\langle 0 | \bar Q \gamma_{\nu} \gamma_5 (i {\overset {\leftrightarrow} {D}}_{\mu_1}) (i {\overset {\leftrightarrow} {D}}_{\mu_2}) Q |\eta_c \rangle = 
i f p_{\nu} T_{\mu_1 \mu_2} + \mbox {higher~order~in~v~terms},
\label{xi}
\eeq
where tensor $T_{\mu_1 \mu_2}$ is defined as
\beq
T_{\mu_1 \mu_2} = g_{\mu_1 \mu_2} - \frac {p_{\mu_1} p_{\mu_2}} {M_{\eta_c}^2}.
\eeq
The constant $f$ can be calculated  contracting the indices $\mu_1$ and $\mu_2$ and regarding the fact that 
time derivative is suppressed
\beq
f =  \frac i {3 M_{\eta_c}} \langle 0| \bar Q \gamma_0 \gamma_5  (i {\bf {\overset {\leftrightarrow} {D} }} )^2 Q |\eta_c \rangle.
\label{eq1}
\eeq
The constant $f$ can also be related to the moment $\langle \xi^2 \rangle$ of the LCWF. To obtain this relation 
let us consider matrix element (\ref{xi}) in the infinite momentum frame ($p \to \infty$). Obviously only 
tensor structure $p_{\mu_1} p_{\mu_2}$ gives contribution to the LCWF of leading twist (\ref{lwf}). 
At the same time at leading twist accuracy matrix element (\ref{xi}) can be found from equation (\ref{xin}). Thus we get
\beq
f_{\eta_c} \langle \xi^2 \rangle = - \frac f {M_{\eta_c}^2}.
\label{eq2}
\eeq
Combining equations (\ref{eq1}), (\ref{eq2}) and the definition of the constant $f_{\eta_c}$ (\ref{def}) we obtain
\beq
\langle \xi^2 \rangle = \frac 1 {3 M_{\eta_c}^2} \frac {\langle 0 |\bar Q \gamma_0 \gamma_5 (i {\bf {\overset {\leftrightarrow} {D}}})^2 Q | \eta_c \rangle} 
{\langle 0 |\bar Q \gamma_0 \gamma_5 Q | \eta_c \rangle}.
\label{xi22}
\eeq
At leading order approximation $M_{\eta_c} = 2 M_c$, 
\beq
\label{4}
\langle 0 |\bar Q \gamma_0 \gamma_5 (i {\bf {\overset {\leftrightarrow} {D}}})^2 Q | \eta_c \rangle &\sim& 4 \langle 0 | \chi^+ (i {\bf {\overset {\leftrightarrow} {D}}} )^2 \psi | \eta_c \rangle, \\ 
\langle 0 |\bar Q \gamma_0 \gamma_5 Q | \eta_c \rangle &\sim& \langle 0 | \chi^+  \psi | \eta_c \rangle,
\eeq
where $\psi$ and $\chi^+$ are Pauli spinor fields that annihilate a quark and an antiquark respectively. 
The factor 4 in equation (\ref{4}) appears since the distance between quark and antiquark in left hand side 
of this equation is $2 {\bf z}$ and ${\bf z}$ in the right hand side. With these relations equation (\ref{xi22}) 
can be rewritten as 
\beq
\langle \xi^2 \rangle = \frac 1 {3 M_c^2} \frac {\langle 0 | \chi^+ (i {\bf {\overset {\leftrightarrow} {D}}})^2 \psi | \eta_c \rangle} 
{\langle 0 | \chi^+  \psi | \eta_c \rangle} = \frac {\langle v^2 \rangle } 3.
\label{xi2}
\eeq
Analogously it is not difficult to calculate similar expressions for higher moments. We will write 
the expressions for $\langle \xi^4 \rangle$ and $\langle \xi^6 \rangle$ without derivation
\beq
\langle \xi^4 \rangle = \frac 1 {5 M_c^4} \frac {\langle 0 | \chi^+ (i {\bf {\overset {\leftrightarrow} {D}}})^4 \psi | \eta_c \rangle} 
{\langle 0 | \chi^+  \psi | \eta_c \rangle} = \frac {\langle v^4 \rangle } 5,
\label{xi4} \\ 
\langle \xi^6 \rangle = \frac 1 {7 M_c^6} \frac {\langle 0 | \chi^+ (i {\bf {\overset {\leftrightarrow} {D}}})^6 \psi | \eta_c \rangle} 
{\langle 0 | \chi^+  \psi | \eta_c \rangle} = \frac {\langle v^6 \rangle } 7. 
\label{xi6}
\eeq
It should be noted here that equations (\ref{xi2})-(\ref{xi6}) were derived within dimensional regularization at
leading order approximation in $\alpha_s$. Moreover, in equations (\ref{xi2})-(\ref{xi6}) the moments are defined
at scale $\mu \sim M_c$ since NRQCD matrix elements $\langle 0 | \chi^+  \psi | \eta_c \rangle, 
\langle 0 | \chi^+ (i {\bf {\overset {\leftrightarrow} {D}}})^2 \psi | \eta_c \rangle, ...$ that will be used in our 
calculations \cite{Bodwin:2006dn} were derived at this scale.

In the framework of NRQCD the matrix elements $\langle 0 | \chi^+ (i {\bf {\overset {\leftrightarrow} {D}} })^n \psi | \eta_c \rangle$ were 
calculated in paper \cite{Bodwin:2006dn} 
\beq
\frac {\langle 0 | \chi^+ ((i {\bf {\overset {\leftrightarrow} {D}}} )^2)^k \psi | \eta_c \rangle} 
{\langle 0 | \chi^+  \psi | \eta_c \rangle} = (m E_B)^k,
\eeq
where $E_B$ is a bound state energy of $\eta_c$ meson, $m$ is the mass of quark in paper \cite{Bodwin:2006dn}.
The main idea of paper \cite{Bodwin:2006dn} is that at leading order approximation in relative velocity NRQCD results 
can be reproduced by potential model with heavy quark potential measured in lattice.
The results of the calculation of the moments  are presented in the fourth column of Table I. 
In the evaluation of the moments we have used the following values of parameters: $M_c = M_{\eta_c} /2$, the mass $m$ and bound 
state energy $E_B$ are taken from paper \cite{Bodwin:2006dn}. The central values of the moments and the errors 
have been calculated according to the approach proposed in the same paper. It should be noted that the error presented in 
Table I is assigned to the uncertainty of potential model parameters used in the calculation. In addition to this
error there is an uncertainty due to higher order $v$ corrections. For the second moment 
one can expect that the error is about $30 \%$. For higher moments this error is larger.

It is seen from Table I that NRQCD predictions for the second and the fourth moments are in good agreement 
with potential model and there is disagreement for the moment $\langle \xi^6 \rangle$ between these two approaches. 
The cause of this disagreement is the fact noted above: due to the large contribution of 
relativistic motion of quark-antiquark pair inside quarkonium it is not possible to apply both 
approaches for higher moments. We believe that both approaches can be used for the estimation 
of the values of the second and the fourth moments. The predictions for the sixth 
and higher moments obtained within both approaches become unreliable.

The aim of this paper is the study of leading twist LCWF of $\eta_c$ meson. But at leading order approximation 
in relative velocity $v$ there is no difference between $\eta_c$ meson and $J/ \Psi$ mesons. So the results 
for the moments obtained within potential models and NRQCD are valid for leading twist LCWFs of $J/ \Psi$ meson. 
Moreover, at leading order approximation in relative velocity the moments of leading twist LCWF of $\eta_c$ meson 
equal to the moments of higher twist LCWFs of $\eta_c$ and $J/\Psi$ mesons since these moments can be 
expresses through the matrix elements: 
$\langle 0 | \chi^+ (i {\bf {\overset {\leftrightarrow} {D} }  } )^2 \psi | \eta_c \rangle$, 
$\langle 0 | \chi^+ (i {\bf {\overset {\leftrightarrow} {D} }  })^4 \psi | \eta_c \rangle$, ...

\begin{table}
$$\begin{array}{|c|c|c|c|c|}
\hline \langle \xi^n \rangle & \mbox{ Buchmuller-Tye } & \mbox{ Cornell } & \mbox{  NRQCD }
& \mbox {QCD }  \\
  & \mbox{ model  \cite{Buchmuller:1980su} } & \mbox{ model \cite{Eichten:1978tg}} & \mbox{\cite{Bodwin:2006dn} } & \mbox{sum rules} \\
\hline
n=2  & 0.086  & 0.084  &  0.075 \pm 0.011   & 0.070 \pm 0.007  \\
\hline
n=4  & 0.020 & 0.019 &   0.010 \pm 0.003  & 0.012 \pm 0.002  \\
\hline
n=6  & 0.0066 & 0.0066 &  0.0017 \pm 0.0007 & 0.0032 \pm 0.0009  \\
\hline
\end{array}$$
\caption{The moments of LCWF obtained within different approaches. In the second and third columns the moments calculated in the framework of 
Buchmuller-Tye and Cornell potential models are presented. In the fourth column NRQCD predictions for the moments are presented. 
In last column the results obtained within QCD sum rules are shown. }
\end{table}

\section{ The moments in the framework of QCD sum rules. }

In the previous sections two approaches to the calculation of the moments of 
leading twist LCWF were considered. The main disadvantage of both approaches is that quarkonium is considered as a nonrelativistic 
bound state of quark-antiquark pair. Due to this the errors of the calculation for both approaches are rather large.
Another approach to the study of the moments of LCWF  based  on QCD sum rules \cite{Shifman:1978bx, Shifman:1978by} was developed by Chernyak and Zhitnitsky
in paper \cite{chernyak}. This approach does not assume that the quarks in 
quarkonium are nonrelativistic. In this section QCD sum rules will be applied to the calculation of the moments of LCWF.

To calculate the moments of LCWF in the framework of QCD sum rules let us consider two-point 
correlator:
\beq
\Pi_n (z,q) = i \int d^4 x e^{i q x} \langle 0| T J_0(x) J_n (0) |0 \rangle = (zq)^{n+2} \Pi_n (q^2), 
\label{cor} \\ \nonumber
J_0 (x) = \bar Q(x) \hat z \gamma_5  Q(x), ~~~ J_n(0) = \bar Q(0) \hat z \gamma_5 (i z^{\sigma} {\overset {\leftrightarrow} {D}_{\sigma}} )^n  Q(0), ~~ z^2=0.
\eeq
It should be noted that explicit dependence of QCD correlator $\Pi_n (q^2)$ on $n$ will be dropped in subsequent equations. 

The calculation of QCD expression for two-point correlator $\Pi_{\rm QCD} (q^2)$ is done through the use of operator product
expansion (OPE) for the T-ordered product of currents. As a result of OPE one obtains besides usual perturbative contribution $\Pi_{\rm pert}(q^2)$ 
also nonperturbative power corrections $\Pi_{\rm npert }(q^2)$, given by QCD vacuum condensates. The diagram that gives 
the leading order perturbative contribution is shown in Fig. 1a.
The leading order nonperturbative power corrections are given by gluon vacuum condensate $\langle G_{\mu \nu} G^{\mu \nu} \rangle$. 
The diagrams that contribute to correlator (\ref{cor}) are shown Fig. 1b-d. The connection to hadrons in the framework of QCD sum rules is 
obtained by matching the resulting QCD expression for current correlator with spectral representation, following the structure of
dispersion relation at $q^2\leq 0$:
\beq
\Pi_{\rm phys} (q^2) &=&  \int_{4 m_c^2}^{\infty} d s \frac { \rho^{\rm phys} (s)  } {s - q^2} + \mbox{ subtractions}, 
\label{phys}
\eeq
\newpage
\begin{figure}[ph]
\begin{picture}(150, 50)
\put(-220,-750){\epsfxsize=21cm \epsfbox{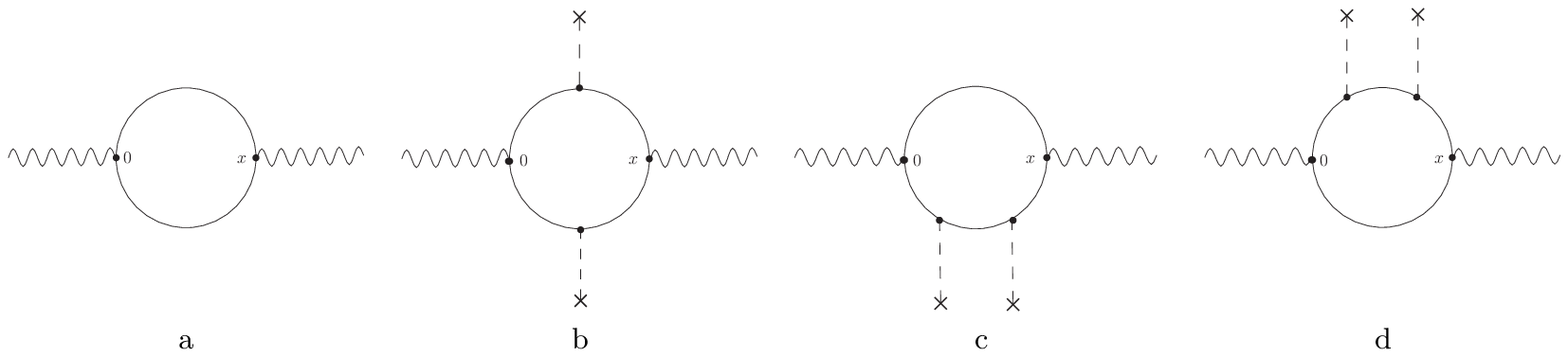}}
\put(-160,-70){{\bf Fig.1:} The diagrams that contribute to the perturbative (a) and nonperturbative (b,c,d) parts of QCD sum rules.}
\end{picture}
\end{figure}
\vspace*{2.5cm}

Assuming that the dispersion relation (\ref{phys}) is well convergent, the physical
spectral functions are generally saturated by the lowest lying hadronic states plus
a continuum starting at some thresholds $s_n$:
\begin{eqnarray}
\rho^{\rm phys} (s) = \rho^{\rm res} (s) +  \theta (s - s_n) \rho^{\rm cont}(s),
\end{eqnarray}
where
\begin{eqnarray}
(qz)^{n+2} \rho^{\rm res} ( q^2 ) = \langle 0 | J_0  | \eta_c \rangle \langle \eta_c | J_n  | 0 \rangle \delta ( q^2 -m^2_{\eta_c}) = f_{\eta_c}^2 (qz)^{n+2} \langle \xi^n \rangle  
\delta (q^2-m^2_{\eta_c}).
\end{eqnarray}
Continuum contribution $\rho^{ \rm cont}$ for $s>s_n$ can be approximated by spectral density of perturbative contribution 
$\sim \mbox {Im} \Pi_{\rm pert}( q^2)$. To get QCD sum rules for the moments of LCWF one should apply operation 
\beq
\Pi^{(m)} (Q^2)= \frac 1 {m!} \biggl ( - \frac {d} {d Q^2}  \biggr )^{m} \Pi (Q^2),
\label{dif}
\eeq
to both QCD and physical expressions for correlator (\ref{cor}) and then equate them. 
Thus we get QCD sum rules for the moments:
\beq
\frac {f_{\eta_c}^2 \langle \xi^n \rangle}  { (m_{\eta}^2+Q^2)^{m+1} } = 
\frac 1 {\pi} \int_{4 m_c^2}^{s_n} ds ~ \frac {\mbox{Im} \Pi_{\rm pert}(s)} {(s+Q^2)^{m+1} }  + \Pi^{(m)}_{\rm npert}(Q^2),
\label{sm}
\eeq
where $\mbox {Im} \Pi_{\rm pert} (s)$, $\Pi^{(m)}_{\rm npert}(Q^2)$ can be written as 
\beq
\mbox {Im} \Pi_{\rm pert} (s) = \frac {3} {8 \pi} v^{n+1}  (\frac 1 {n+1} - \frac {v^2} {n+3} ), ~~~~ v^2 = 1 - \frac {4 m_c^2} {s},
\label{pert}
\eeq
\beq
\label{power}
\Pi^{(m)}_{\rm npert}(Q^2) &=& \Pi_1^{(m)} (Q^2) + \Pi_2^{(m)} (Q^2)+\Pi_3^{(m)} (Q^2), \\ \nonumber
\Pi_1^{(m)} (Q^2) &=& \frac {\langle \alpha_s G^2 \rangle} {24 \pi }
(m+1) \int_{-1}^1 d \xi~  \biggl (\xi^n + \frac {n(n-1)} 4 \xi^{n-2} (1- \xi^2) \biggr )
\frac {(1- \xi^2)^{m+2}} {\bigl ( 4 m_c^2  + Q^2 (1- \xi^2 ) \bigr )^{m+2} }, \\ \nonumber
\Pi_2^{(m)} (Q^2) &=& - \frac {\langle \alpha_s G^2 \rangle} {6 \pi} m_c^2 ( m^2+ 3m +2)
\int_{-1}^1 d \xi~ 
  \xi^n \bigl ( 1+ 3 \xi^2 \bigr ) \frac {(1- \xi^2)^{m+1}} {\bigl ( 4 m_c^2  + Q^2 (1- \xi^2 ) \bigr )^{m+3} }, \\ \nonumber
\Pi_3^{(m)} (Q^2) &=&  \frac {\langle \alpha_s G^2 \rangle} {384 \pi }
(n^2-n) (m+1) \int_{-1}^1 d \xi~  \xi^{n-2} 
\frac {(1- \xi^2)^{m+3}} {\bigl ( 4 m_c^2  + Q^2 (1- \xi^2 ) \bigr )^{m+2} },
\eeq
here $Q^2=-q^2$. In the original paper \cite{Shifman:1978by} 
the method QCD sum rules was applied at $Q^2=0$. However as it was shown in paper \cite{Reinders:1984sr}
there is large contribution of higher dimensional operators at $Q^2=0$ which grows rapidly with $m$. 
To suppress this contribution in our paper sum rules (\ref{sm}) will be applied at $Q^2=4 m_c^2$.

In the numerical analysis of QCD sum rules the values of parameters $m_c$ and $\langle \alpha_s G^2/ \pi \rangle$ will be 
taken from paper: \cite{Reinders:1984sr}
\beq
m_c = 1.24 \pm 0.02 ~\mbox {GeV}, ~~ \langle \frac {\alpha_s} {\pi} G^2 \rangle = 0.012 \pm 30 \% ~\mbox {GeV}^4.
\label{param}
\eeq
First sum rules (\ref{sm}) will be applied to the calculation of the constant $f_{\eta_c}^2$. 
It is not difficult to express the constant $f_{\eta_c}^2$  from equation (\ref{sm}) at $n=0$ as a function
of $m$. For too small values of $m$ ($m<m_1$) there is large contributions from  higher resonances 
and continuum which spoil sum rules (\ref{sm}). Although for $m \gg m_1$ these contributions
are strongly suppressed, it is not possible to apply sum rules for too large $m$ ($m>m_2$) 
since the contribution arising from higher dimensional vacuum condesates rapidly grows with $m$ what invalidates 
our approximation. If $m_1<m_2$ there is some region of applicability of sum rules (\ref{sm}) $[m_1, m_2]$ where the resonance and 
the higher dimensional vacuum condensates contributions are not too large. Within this region 
$f_{\eta_c}^2$ as a function of $m$ varies slowly and one can determine the value of this constant.
The value of the continuum threshold $s_n$ must be 
taken so that to appear stability region $[m_1, m_2]$. Our calculation shows that for 
central values of the parameters (\ref{param}) there is stability in the region $m>4$ if 
one takes $s_n>3.6^2$ GeV$^2$. In paper \cite{Vainshtein:1978nr}, where the mass of $\eta_c$ meson was studied within 
QCD sum rules, the authors took $s_n= \infty$. In our paper we will use the same value of the threshold. 
Applying approach described above we get 
\beq
f_{\eta_c}^2 = 0.120 \pm 0.002 \pm 0.005 \pm 0.016 ~ \mbox {GeV}^2.
\label{feta}
\eeq
The region of stability in the determination of the constant $f^2_{\eta_c}$ 
begins at $m_1=6$ and ends at $m_2=10$. The first error in (\ref{feta}) corresponds the variation of $f_{\eta_c}^2$
within this region. The second and the third errors in (\ref{feta}) correspond to the 
variation  of the gluon condensate $\langle \alpha_s G^2 \rangle$ and the mass $m_c$ within ranges (\ref{param}). From the 
results (\ref{feta}) one sees that the main error in determination of the constant $f_{\eta_c}^2$ 
results from the variation of the parameter $m_c$. This fact represents well known property: 
high sensitivity of QCD sum rules to the mass parameter $m_c$.

Next let us consider the second moment of LCWF $\langle \xi^2 \rangle$ in the framework of QCD sum rules. One way 
to find the value of $\langle \xi^2 \rangle$ is to determine the value of $f_{\eta}^2 \langle \xi^2 \rangle$ 
from sum rules (\ref{sm}) at $n=2$ and then extract $\langle \xi^2 \rangle$. 
However, as it was noted above, this approach suffers from high sensitivity of right side of equation (\ref{sm}) to the variation 
of the parameter $m_c$. Moreover, the quantity $f_{\eta}^2 \langle \xi^2 \rangle$ includes not only the error 
in determination of  $\langle \xi^2 \rangle$, but also the error in $f_{\eta}^2$. To remove this disadvantages 
in our calculation we will consider the ratio of sum rules at $n=2$ and $n=0$: $f_{\eta}^2 \langle \xi^2 \rangle / f_{\eta}^2$.
The moments $\langle \xi^4 \rangle, \langle \xi^6 \rangle$ will be considered analogously. Applying
standard procedure one gets the results:
\beq
\label{res}
\langle \xi^2 \rangle &=& ~0.070 \pm 0.002 \pm 0.007 \pm 0.003,  \\
\langle \xi^4 \rangle &=& ~0.012 \pm 0.001 \pm 0.002 \pm 0.001, \nonumber \\
\langle \xi^6 \rangle &=& ~0.0032 \pm 0.0002 \pm 0.0009 \pm 0.0003. \nonumber  
\eeq
In the calculation of moments the stability was achieved within the interval $m \in [9,13]$. 
The first error in (\ref{res}) corresponds the variation 
within the region of stability. The second and the third errors in (\ref{res}) correspond to the 
variation  of the gluon condensate $\langle \alpha_s G^2 \rangle$ and the mass $m_c$ within ranges (\ref{param}). It is 
seen that, as one expected, the sensitivity of the ratio $f_{\eta}^2 \langle \xi^n \rangle / f_{\eta}^2$ to the variation of 
$m_c$ is rather low. The main source of uncertainty is the variation of gluon condensate $\langle \alpha_s G^2 \rangle$.
In the fourth column of Table I the results of our calculation are presented. The errors in Table I correspond 
to the main source of uncertainty --- the variation of gluon condensate $\langle \alpha_s G^2 \rangle$.

It is not difficult to show that in the calculation of QCD expression for the correlator (\ref{cor}) 
characteristic virtuality of quark is $\sim (4 m_c^2 + Q^2)/m \sim m_c^2$. So the values of the 
moments (\ref{res}) are defined at scale $\sim m_c^2$. It should be noted here that, in all approaches
applied to the calculation of the moments, the quantities $\langle \xi^n \rangle$ are defined at different scales. 
But the scale dependence appears at NLO $\alpha_s$ correction and since all these scales are of order 
of $\sim m_c$, the variation of the scale near $\sim m_c$ will not change our results considerably. For this reason we believe that
it is possible to compare the results obtained within different approaches. 

From Table I it is seen that the larger the number of the moment $n$ the larger the 
uncertainty due to the variation of vacuum gluon condensate. This property is a consequence 
of the fact that the role of power corrections in the sum rules (\ref{sm}) grows with $n$.
From this one can conclude that there is considerable nonperturbative contribution to 
the moments $\langle \xi^n \rangle $ with large $n$ what means that nonperturbative effects are very 
important in relativistic motion of quark-antiquark pair inside the meson. The second 
important contribution to QCD sum rules (\ref{sm}) at large $n$ is QCD radiative corrections
to perturbative part $\Pi_{\rm pert}(Q^2)$.
Unfortunately today one does not know the expression for these corrections and for 
this reason we cannot include them to sum sules (\ref{sm}). We can only say 
that these corrections grow with $n$ and probably the size of radiative corrections to 
the ratio $f_{\eta}^2 \langle \xi^n \rangle / f_{\eta}^2$ is not too big for not too large $n$. 
Thus one can expect that QCD radiative corrections will not change dramatically our results 
for the moments  $n=2$ and $n=4$. But the radiative corrections to  $\langle \xi^6 \rangle$ may be important. 

It is interesting to estimate relative velocity of quark-antiquark pair inside $\eta_c$ meson using the results 
obtained within QCD sum rules. In the section devoted to the calculation of the moments in the framework 
of NRQCD we have derived formula (\ref{xi2}). At leading order approximation this formula allows one to connect the square of 
relative velocity and the second moment of LCWF. Applying this formula one gets
\beq
\langle v^2 \rangle_{\eta_c} = 3 \langle \xi^2 \rangle_{\eta_c} = 0.21 \pm 0.02.
\label{vv}
\eeq
This value is in good agreement with potential models estimations of $\langle v^2 \rangle$ \cite{Buchmuller:1980su, Eichten:1979ms}. The error in (\ref{vv}) 
corresponds to the main source of uncertainty of our results --- the variation of gluon condensate $\langle \alpha_s G^2 \rangle$.
In addition to this error there is an uncertainty due to higher order $v$ corrections $\sim 30 \%$.

\newpage
\begin{figure}[ph]
\begin{picture}(150, 50)
\put(-130,-500){\epsfxsize=15cm \epsfbox{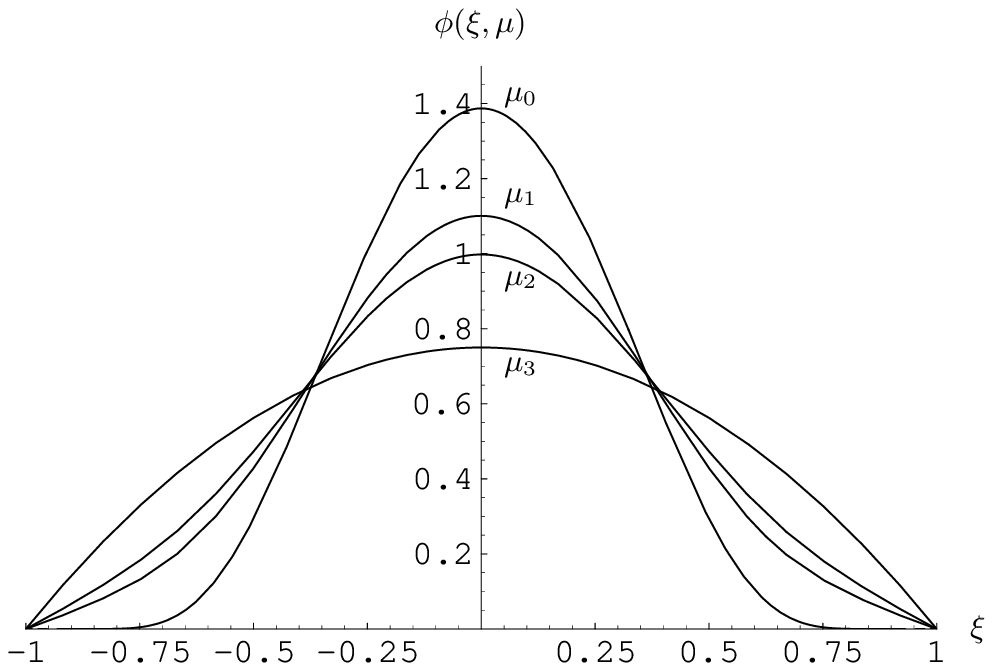}}
\put(-100, -100){{\bf Fig.2:} The LCWF (\ref{model}) at scales $\mu_0 =1.2~\mbox{GeV} , \mu_1 = 10 ~\mbox{GeV}, \mu_2 =100 ~\mbox{GeV}, \mu_3 = \infty$.}
\end{picture}
\end{figure}
\vspace*{3.5cm}

It should be noted here that in our calculation we have used the 
value of the threshold $s_n=\infty$. But if one takes the value of the threshold 
energy within a reasonable interval this will lead to the shift of the stability region 
to lower $m$ and the results of our calculation will be changed not greater than by 
few percents for $\langle \xi^2 \rangle, \langle \xi^4 \rangle$ and $10 \%$ for $\langle \xi^6 \rangle$.

We believe that the values of the moments (\ref{res}) calculated within QCD sum rules are 
more reliable than the ones obtained in the framework of potential models and NRQCD. For this reason 
we consider 
results (\ref{res}) as the main result of our paper and below 
these values will be used.

\section{The model for the LCWF of $\eta_c$ meson.}

To build the model of the leading twist LCWF of $\eta_c$ meson let us
write Borel version \cite{Shifman:1978bx, Shifman:1978by} of sum rules (\ref{sm}) without continuum contribution $\rho^{\rm cont}$ and power corrections $\Pi_{\rm npert}$
\beq
f_{\eta_c}^2 \langle \xi^n \rangle e^{-  {m_{\eta_c}^2}/ {M^2}} = 
\frac {M^2} {4 \pi^2} \int_{-1}^1 d \xi ~\xi^n ~\frac 3 4 (1- \xi^2) \mbox{exp} \biggl (  - \frac {4 m_c^2} {M^2} \frac 1 {1-\xi^2} \biggr ).
\label{smn}
\eeq
Within this approximation the LCWF can be written in the form
\beq
\phi(\xi, \mu \sim m_c) = c( \beta ) (1- \xi^2) \mbox{exp} \biggl (  -  \frac {\beta} {1-\xi^2} \biggr ),
\label{model}
\eeq
where $ c (\beta)$ is a normalization constant and $\beta$ is a constant that will be calculated below.
Although sum rules (\ref{smn}) is much simpler that sum rules (\ref{sm}), to reproduce the results obtained in the 
previous section  with good accuracy it is sufficient to use LCWF in  the form of equation (\ref{model}).
To fix the constant $\beta$ the moment $\langle \xi^2 \rangle$ will be taken. Moreover 
we will suppose that the results (\ref{res}) obtained in the previous section are defined at scale $\mu_0=1.2~ \mbox{GeV} \sim m_c$.
Thus we get $\beta = 3.8 \pm 0.7$. The constant $c(\beta)$ can be determined from normalization condition (\ref{norm}).  It is not difficult to see that $1/ \beta \sim v^2$ as one can expect. 
The moments of this wave function are
\beq
\langle \xi^2 \rangle &=& ~0.070 \pm 0.007,  \\
\langle \xi^4 \rangle &=& ~0.012 \pm 0.002, \nonumber \\
\langle \xi^6 \rangle &=& 0.0030 \pm 0.0009. \nonumber  
\eeq
At central value $\beta=3.8$ the constant $c(\beta) \simeq 62$. Below the LCWF (\ref{model}) 
at central value will be used. 

LCWF (\ref{model}) is defined at scale $\mu=\mu_0$.  It is not difficult to 
calculate this function at any scale $\mu > \mu_0$ using conformal expansion (\ref{conf_exp}). 
The LCWFs at scales $\mu_0 =1.2~\mbox{GeV} , \mu_1 = 10 ~\mbox{GeV}, \mu_2 =100 ~\mbox{GeV}, \mu_3 = \infty$
are shown in Fig. 2. The moments of LCWFs at scales $\mu_0 = 1.2~\mbox{GeV}, \mu_1 = 10 ~\mbox{GeV}, \mu_2 =100 ~\mbox{GeV}, \mu_3 = \infty$
are presented in second, third, fourth and fifth columns of Table II.

The model (\ref{model}) has some interesting properties. 
From conformal expansion (\ref{conf_exp}) one can derive the expressions that determine the evolution of the moments:
\beq
\label{xxi}
\langle \xi^2 \rangle_{\mu} &=& \frac 1 5 + a_2(\mu) \frac {12} {35}, \\ \nonumber
\langle \xi^4 \rangle_{\mu} &=& \frac 3 {35} + a_2(\mu) \frac {8} {35} + a_4(\mu) \frac {8} {77}, \\ \nonumber
\langle \xi^6 \rangle_{\mu} &=& \frac 1 {21} + a_2(\mu) \frac {12} {77} + a_4(\mu) \frac {120} {1001} + a_6(\mu) \frac {64} {2145}.
\eeq
Similar relations can be found for any moment. Further let us consider the expression for the second moment $\langle \xi^2 \rangle$.
In our paper we have found the value $\langle \xi^2 \rangle$
with some error at scale $\mu = \mu_0$. This means that the value of the coefficient $a_2( \mu = \mu_0)$ 
was found with some error. The coefficient $a_2$ decreases as scale increases.
So the error in $a_2$ and consequently in $\langle \xi^2 \rangle$ decreases as scale increases. At 
infinitely large scale there is no error in $\langle \xi^2 \rangle$ at all. Our calculations show 
that the error $10 \%$ in $\langle \xi^2 \rangle$ at scale $\mu=\mu_0$ decreases to $4 \%$ at scale
$\mu = 10$ GeV. Applying relations (\ref{xxi}) it is not difficult to show that similar improvement of the 
accuracy happens for higher moments. The improvement of the 
accuracy allows us to expect that the model (\ref{model}) at larger scales will be rather good even if QCD
radiative corrections to the results (\ref{res}) are large.

From Fig. 2 one sees that LCWF at scale $\mu=\mu_0$ practically vanishes 
in the regions $0.75 < |\xi| < 1$. In this region the motion of quark-antiquark pair is relativistic 
and vanishing of LCWF in this region means that at scale $\mu = \mu_0$ charmonium can be considered 
as a nonrelativistic bound state of quark-antiquark pair with characteristic velocity $v^2 \sim 1/ \beta \sim 0.3$.
Further let us regard the function $\phi(\xi, \mu=\mu_0)$ as a conformal expansion (\ref{conf_exp}). 
To get considerable suppression of the LCWF in the region $0.75 < |\xi| < 1$ one should require 
fine tuning of the coefficients of conformal expansion $a_n ( \mu=\mu_0 )$. The evolution of the constants $a_n$( especially with large $n$)
near $\mu = \mu_0$ is rather rapid ( see formulas (\ref{ren}) and (\ref{an_dim})) and if there is fine tuning of the constants
at scale $\mu=\mu_0$ this fine tuning will be rapidly broken at larger scales. This property is well seen 
in Table II and Fig. 2. From Fig. 2 it is seen that there is relativistic tail in the region $0.75 < |\xi| < 1$
for scales $\mu= 10, 100$ GeV which is absent at scale $\mu=\mu_0$. Evidently this tail cannot be regarded
in the framework of NRQCD. This means that, strictly speaking, at some scale 
charmonium can not be considered as nonrelativistic particle and the application of NRQCD to the production of charmonium at large scales
may lead to large error. Although in our arguments we have used 
the model of LCWF (\ref{model}) it is not difficult to understand that 
the main conclusion is model independent.

According to the velocity scaling rule \cite{Bodwin:1994jh} the moments $\langle \xi^n \rangle$ of LCWF 
depend on relative velocity as $\sim v^n$. It is not difficult to show that the moments of LCWF (\ref{model})
satisfy these rules. Now let us consider the expressions that allows one to connect 
the coefficients of conformal expansion $a_n$ with the moments $\langle \xi^n \rangle$.
These expressions for the moments $\langle \xi^2 \rangle, \langle \xi^4 \rangle, \langle \xi^6 \rangle$
are given by formulas (\ref{xxi}). It causes no difficulties to find similar 
expressions for any moment. From expressions (\ref{xxi}) one sees that 
to get velocity scaling rules: $\langle \xi^n \rangle \sim v^n$ at some scale one should require 
fine tuning of the coefficients $a_n$ at this scale. But, as was already noted above, 
if there is fine tuning of the coefficients at some scale this fine tuning will be 
broken at larger scales. From this one can conclude that velocity scaling rules 
are broken at large scales. 

Consider the moments of LCWF (\ref{model}) at infinite scale. It is not difficult to find that
\beq
\langle \xi^n \rangle_{\mu=\infty} = \frac 3 {(n+1) (n+3)}.  
\eeq
From last equation one can find that $\langle \xi^n \rangle $ does not scale as $v^n$ as 
velocity scale rules \cite{Bodwin:1994jh} require. Thus scaling rules obtained in paper \cite{Bodwin:1994jh}
are broken for asymptotic function. Actually one does not need to set the scale $\mu$ to infinity to break 
these rules. For any scale $\mu > \mu_0$ there is a number $n_0$ for which the moments $\langle \xi^n \rangle,~ n>n_0$
violate velocity scaling rules. This property is a consequence of the following fact:
beginning from some $n=n_0$ the contribution of the relativistic tail of LCWF, that appears at scales $\mu>\mu_0$, 
to the moments becomes considerable. 

The amplitude $T$ of any hard process with charmonium meson production can be written as a convolution 
of LCWF $\Phi( \xi)$ with hard kernel $H(\xi)$ of the process. If one expands this kernel over $\xi$ and substitute 
this expansion to the amplitude $T$ one gets the results:
\beq
T = \int d \xi H(\xi) \Phi(\xi) = \sum_{n} \frac {H^{(n)} (0)}  {n!} \langle \xi^n \rangle.
\label{s}
\eeq
If one takes the scale $\mu \sim \mu_0$ in formula (\ref{s}) than moments $\langle \xi^n \rangle$ 
scale according to the velocity scaling rules $\sim v^n$ and  one gets usual NRQCD expansion of the amplitude.
However due to the presence of the scale of the hard process $\mu_h \gg \mu_0$ there appears large 
logarithms $\log {\mu_h/\mu_0}$ which spoil NRQCD expansion (\ref{s}). To remove this 
large logarithms one should take $\mu \sim \mu_h$. But at large scales velocity scaling rules 
are broken and application of NRQCD is questionable.

\begin{table}
$$\begin{array}{|c|c|c|c|c|c|c|}
\hline \langle \xi^n \rangle & \phi(\xi, \mu_0=1.2~\mbox{GeV}) & \phi(\xi, \mu_1=10~\mbox{GeV}) & \phi(\xi, \mu_2=100~\mbox{GeV})
& \phi(\xi, \mu_3=\infty ) &  \mbox{BC \cite{Bondar:2004sv}} &  \mbox{BKL \cite{Bodwin:2006dm}} \\
  \hline
n=2  &  0.070 & 0.12  & 0.14 &  0.20 & 0.13 & 0.019 \\
\hline
n=4  & 0.012 &  0.040 & 0.052 & 0.086 &  0.040 & 0.0083 \\
\hline
n=6  & 0.0032 & 0.019 & 0.026 & 0.048 & 0.018 & 0.0026 \\
\hline
\end{array}$$
\caption{ The moments of LCWF (\ref{model}) proposed in this paper at scales $\mu_0 = 1.2~\mbox{GeV}, \mu_1 = 10 ~\mbox{GeV}, \mu_2 =100 ~\mbox{GeV}, \mu_3 = \infty$
are presented in second, third, fourth and fifth columns. The moments of LCWF proposed in paper \cite{Bondar:2004sv} are shown 
in the sixth column. The moments of LCWF proposed in paper \cite{Bodwin:2006dm} are shown 
in the sixth column. }
\end{table}

\section{Comparison with other models.}

{\bf Asymptotic wave function.}

We have already noted that if in conformal expansion (\ref{conf_exp}) the scale $\mu$ tends to infinity 
LCWF tends to the function 
\beq
\phi_{as} = \frac 3 4 (1- \xi^2),
\label{as}
\eeq
which is called asymptotic. This wave function does not depend on the mesons with the same quantum 
numbers. For instance, there is one LCWF defined by equation (\ref{as}) for $\eta_c$ and $\pi$ mesons. 
From this point of view the distinction of quarkonium LCWF from it's asymptotic form can be considered
as a measure of relativism of the quarkonium. For particularly nonrelativistic object the relation 
$\langle \xi^2 \rangle \ll \langle \xi^2 \rangle_{\mu=\infty}$ must be satisfied.

The moments of asymptotic LCWF are presented in the fifth column of Table II. One sees that 
although at scale $\mu=\mu_0$ $\eta_c$ can be considered as nonrelativistic object with $v^2 \sim 0.3$, 
the relation for the particularly nonrelativistic object $\langle \xi^2 \rangle_{\mu=\infty} \gg \langle \xi^2 \rangle_{\mu=m_c}$ 
is badly satisfied. At scale $\mu=10$ GeV this relation is not satisfied at all and at this scale 
$\eta_c$ can not be considered as a nonrelativistic object.
This fact has already been noted in the previous section.

{\bf The LCWF proposed by Bondar and Chernyak in paper \cite{Bondar:2004sv}. }

To resolve contradiction between NRQCD prediction  \cite{Braaten:2002fi}
and experimental measurements  of the cross section of the process $e^+ e^- \to J/ \Psi \eta_c$
at energy $\sqrt s = 10.6$ GeV \cite{Abe:2004ww, Aubert:2005tj}
Bondar and Chernyak proposed the following model of LCWF \cite{Bondar:2004sv}:
\beq
\phi_{BC} (\xi )=c (v^2)\,\phi_{as}(\xi)  \Biggl \{ \frac{x_1 x_2}
{[1-4x_1x_2(1-v^2)]} \Biggr \}^{1-v^2}\, ,
\label{BC}
\eeq
where $v^2=0.3$ is relative velocity of quark-antiquark pair in $\eta_c$ meson, $c(v^2)$ is 
the coefficient which is fixed by the wave function normalization $\int d x \phi_{BC}(\xi) =1$, $\phi_{as}(\xi)$
is the asymptotic LCWF (\ref{as}). It should be noted that LCWF (\ref{BC}) breaks 
velocity scaling rules. 

The moments of this wave function are presented in the sixth column of Table II.
According to paper \cite{Bondar:2004sv} LCWF (\ref{BC}) is defined at scale $\mu \sim \mu_0$. 
However from Table II it can be seen that the second moment of this function is approximately two times larger than
the second moments calculated in this paper. For higher moments the difference is greater. The moments of this wave function are in better
agreement with the moments of $\phi(\xi, \mu=10 ~\mbox{GeV})$. Our calculation shows that 
within the region $\xi \in  (-0.8,0.8)$ the function $\phi(\xi, \mu=10 ~\mbox{GeV})$ differs 
from function (\ref{BC}) by less than $15 \%$. In the end point region $| \xi | \sim 1$
these functions considerably differ from each other but this region does not give main 
contribution to the cross section of the process $e^+ e^- \to J/ \Psi \eta_c$. 
The fact that the scale $\mu =10$ GeV is an energy at which  the cross section of the process $e^+ e^- \to J/ \Psi \eta_c$ was 
measured \cite{Abe:2004ww, Aubert:2005tj} allows us to understand why application of LCWF (\ref{BC})
led to a good agreement with the experiments.

{\bf Effective LCWF proposed by Bodwin, Kang and Lee (BKL) in paper \cite{Bodwin:2006dm}.}

To reconcile light cone and NRQCD approaches to calculating $e^+ e^- \to J/ \Psi + \eta_c$ the 
authors of paper \cite{Bodwin:2006dm} proposed LCWF of $\eta_c$ meson. In our paper 
we are not going to discuss  the approach applied in paper \cite{Bodwin:2006dm}. Only main results of this paper 
will be considered. Thus LCWF obtained in this paper is
\beq
\phi_{0} (\xi)  \sim \int_{\sqrt {d( \xi )} }^{\infty} dp  ~\sqrt{ p^2 + m_c^2 } \int_{0}^{\infty} dr~ \psi(r)~ \sin {(pr)}, 
\label{bkl0}
\eeq
where 
\beq
d(\xi) = m_c^2 \frac {\xi^2} {1- \xi^2},
\eeq
$m_c$ is the pole mass of $c$ quark, $\psi(r)$ is a solution of Schrodinger equation with 
the potential $-k/r + r/a^2$. In our calculation potential model parameters and pole mass $m_c$ are taken from paper 
\cite{Bodwin:2006dm}. LCWF (\ref{bkl0}) will not be discussed here since LCWF (\ref{bkl0}) and (\ref{BC}) 
are similar in shape and have approximately equal moments. It should be noted here only that LCWF (\ref{bkl0}) is wider than model of LCWF (\ref{model})
proposed in this paper.

Expression (\ref{bkl0}) contains high relative  momentum-$p$ tail which cannot be regarded within 
nonrelativistic potential model. To resolve this problem the authors  of paper \cite{Bodwin:2006dm} introduced effective LCWF that removed 
this high  momentum tail from LCWF (\ref{bkl0}). 
\newpage
\begin{figure}[ph]
\begin{picture}(150, 50)
\put(-20,-80){\epsfxsize=6cm \epsfbox{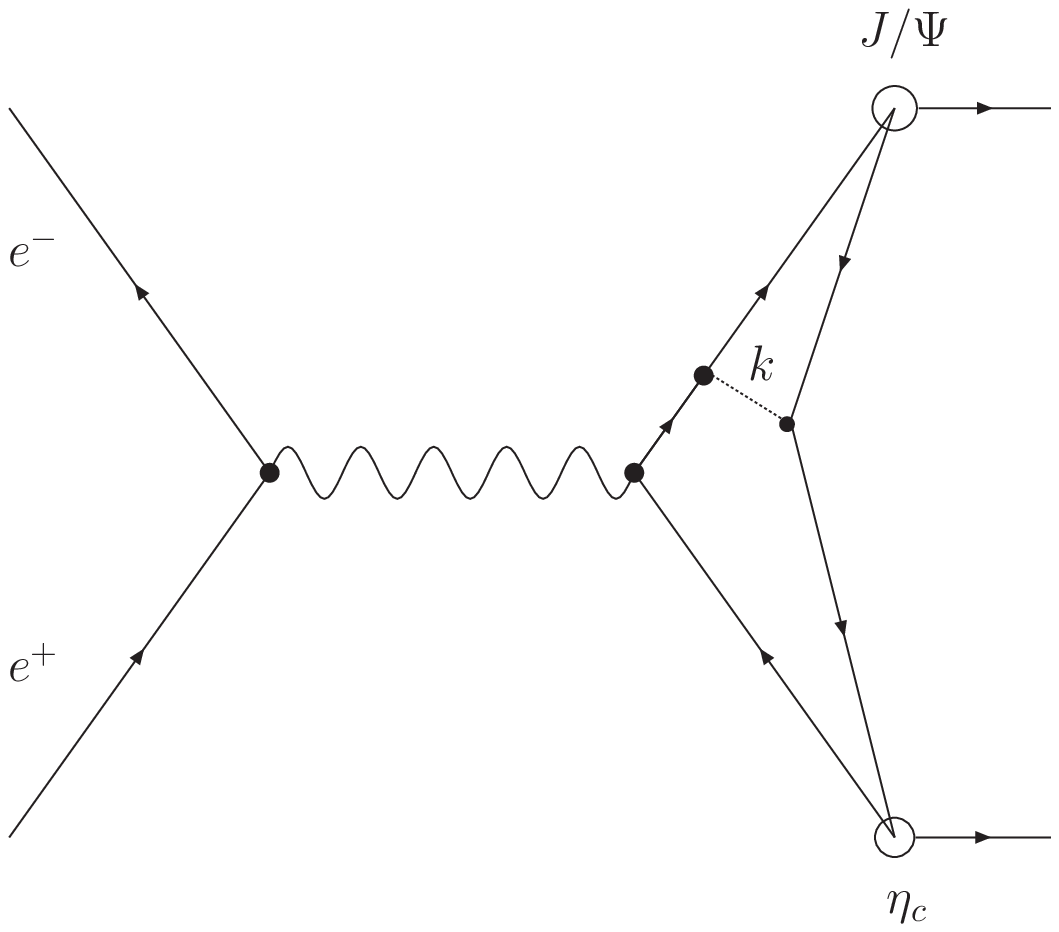}}
\put(-110, -90){{\bf Fig.3:} The diagram that gives contribution to the amplitude of the process $e^+ e^- \to J/\Psi \eta_c$}
\end{picture}
\end{figure}
\vspace*{3.5cm}

The resulting effective LCWF is
\beq
\phi_{BKL}( \xi) = \phi_0( \xi) - \Delta \phi( \xi),
\label{bkl}
\eeq
where 
\beq
\Delta \phi( \xi) = \frac {2 \alpha_s } {3 \pi} \frac {m } {m_c} \biggl [ \frac {\sqrt{1 - \xi^2}} {\xi^2} + 
\frac 1 2 \log { \frac {1+\sqrt{1-\xi^2}} {1-\sqrt{1-\xi^2}} }  \biggr ]_{+\xi}.
\label{delta}
\eeq
In the last equation the following designations were used: $m$ and $\alpha_s$ is the mass of $c$ quark and coupling constant in potential model \cite{Bodwin:2006dm} and
$[f(\xi)]_{+\xi}$ means
\beq
\int_{-1}^{1} d \xi [f(\xi)]_{+ \xi} H(\xi)= \int_{-1}^{1} d \xi f(\xi) [H(\xi)-H(0)].
\eeq
As it was noted above  effective function (\ref{bkl}) is not true LCWF. 
The subtraction $\Delta \phi(\xi)$ is a formal device to remove the
high-momentum contributions that are not calculated reliably in a
potential model and to avoid double count the NRQCD order-$\alpha_s$
corrections. However, we believe that one can compare the 
LCWF (\ref{model}) and function (\ref{bkl}) for the following reasons. 
First, as it was noted above at scale $\mu \sim \mu_0$ there is no high-momentum contribution in model (\ref{model}).
Moreover, the calculation of the moments of LCWF was done within QCD sum rules at leading order approximation in $\alpha_s$, 
so there is no double counting in the framework of light cone formalism\footnote{Note, however, that the NRQCD-formalism order-$\alpha_s$
corrections are different from the light cone formalism order-$\alpha_s$
corrections.}. Second, the main goal of this comparison is to understand 
what results one can expect if instead of LCWF (\ref{model})  function (\ref{bkl}) will be applied. 

The moments of effective LCWF (\ref{bkl}) are presented in the seventh column of Table II. From 
this table it is seen that the moments of model (\ref{bkl}) are much smaller not only than the predictions 
obtained within potential models and QCD sum rules but also than NRQCD 
predictions for the moments. 

In model (\ref{bkl}) large momentum tail which can not be calculated reliably in the framework 
of potential model is removed. Due to this the contribution of relativistic motion of quark-antiquark pair 
inside the $\eta_c$ meson to LCWF becomes strongly suppressed. As we have already noted from the point of conformal expansion 
(\ref{conf_exp}) this suppression can be achieved by fine tuning of the coefficients $a_n$.
But if this fine tuning takes place at some scale ( in paper \cite{Bodwin:2006dm} this happens at scale $\mu \sim \mu_0$)
it will be broken at larger scales where effective LCWF (\ref{bkl}) is applied. 

To illustrate this point let us consider the calculation of the cross section of the process $e^+ e^- \to J/ \Psi + \eta_c$  at energy $\sqrt s =10.6$ GeV
with the LCWF proposed in our paper. Unfortunately to calculate the amplitude of the process $e^+ e^- \to J/ \Psi + \eta_c$ 
one must know not only leading twist wave functions but also next-to-leading twist wave functions of $\eta_c$ and $J/\Psi$ 
mesons (see papers \cite{Bondar:2004sv, Braguta:2005kr}). In our calculation the following model for LCWF of $\eta_c$ and $J/\Psi$  
mesons will be used:
\beq
\phi_i (\xi, \mu) \sim \phi_i^{as} (\xi) \biggl (  \frac { \phi( \xi, \mu) } {1- \xi^2} \biggr ),
\label{mod}
\eeq
where $\phi_i^{as}(\xi)$ is the asymptotic LCWF that corresponds to the function $\phi_i (\xi, \mu)$,
the function  $\phi( \xi, \mu)$ was defined by equation (\ref{model}). One can expect that at scale $\mu \sim \mu_0$ this model
is rather good for LCWFs for $\eta_c$ and $J/\Psi$ mesons since at this scale $\eta_c$ and $J/\Psi$ mesons 
have equal moments up to the corrections of higher order in relative velocity expansion. At larger 
scales the model is not so good since the evolutions of 
LCWFs $\phi_i(\xi, \mu)$ with $\mu$ are different. But we believe that it is sufficient to apply this 
model to the estimation of the cross section  of the process $e^+ e^- \to J/ \Psi + \eta_c$. 
It should be noted here that in model (\ref{mod}) large relative momentum tail of the LCWFs at scale 
$\mu \sim \mu_0$ is strongly suppressed as it was required in paper \cite{Bodwin:2006dm}. The calculation of the cross 
section of the process $e^+ e^- \to J/ \Psi + \eta_c$ \cite{Braguta:2005kr} 
with the wave function taken at scale $\mu \sim \mu_0$ gives the result
\beq
\sigma(e^+ e^- \to J/ \Psi + \eta_c) = 12.1 ~\mbox{fb.}
\eeq
This result is larger than $\sigma = 8.19$ fb obtained in paper \cite{Bodwin:2006dm} since LCWF (\ref{model})
is wider than (\ref{bkl}).  

If the same calculation is done with LCWF taken at scale approximately equal to the momentum running through gluon  
propagator (see Fig. 3) $\sqrt {k^2} = \sqrt s /2 = 5$ GeV than we have
\beq
\sigma(e^+ e^- \to J/ \Psi + \eta_c) = 25.1 ~\mbox{fb.}
\eeq

{\bf LCWF proposed by Ma and Si(MS) in paper \cite{Ma:2006hc}. }

Last light cone distribution amplitude that will be considered in our paper is the LCWF proposed in paper \cite{Ma:2006hc}. 
The authors of this paper build LCWF as a series in $\alpha_s$
\beq
\phi_{BS}(\xi, \mu) = \phi^0 (\xi) + \phi^1 (\xi, \mu) + O( \alpha_s^2).
\label{ms}
\eeq
At leading order approximation the function $\phi^0 \sim \delta (\xi)$ was taken. Next-to-leading function $\phi^1(\xi, \mu)$
can be found in paper \cite{Ma:2006hc}. Due to the singular function $\phi^1 (\xi, \mu)$ and infinitely narrow $\phi_0(\xi)$
at scale $\mu \sim \mu_0$ the whole LCWF (\ref{ms}) has negative moments
\beq
\langle \xi^n \rangle_{MS} = \int_{-1}^1 d \xi~ \xi^n  \phi_{MS}(\xi, \mu = m_c) <0. 
\eeq
For this reason we do not show the moments of this LCWF in Table II. It should be noted here
that as it was shown above LCWF of $\eta_c$-meson is rather wide and we don't think that it is possible to consider 
it as an infinitely narrow object at leading order approximation. 

\section{Conclusion.}

In this paper we have calculated the moments of leading twist light cone wave function (LCWF) of $\eta_c$ mesons
within three approaches. In the first approach we have applied Buchmuller-Tye and Cornell potential models
to the calculation of the moments of LCWF. In the second approach the moments of LCWF were calculated in the 
framework of NRQCD. In the third approach the method QCD sum rules was applied to the calculation of the moments. 
The results obtained within these three approaches are in good agreement with each other for the second 
moment $\langle \xi^2 \rangle$. There is a little disagreement between the predictions for the fourth 
moment $\langle \xi^4 \rangle$. The disagreement between the approaches becomes dramatic for the 
sixth moment $\langle \xi^6 \rangle$. The cause of this disagreement consists in the considerable 
contribution of relativistic motion of quark-antiquark pair inside $\eta_c$ meson to higher moments
which cannot be regarded reliably in the framework of potential models and NRQCD. The approach based on
QCD sum rules is more reliable, especially for higher moments since it does not consider $\eta_c$-meson 
as a nonrelativistic object. The main problem of QCD sum rules is that since there is no expressions 
of radiative corrections to sum rules one does not know the size these corrections. 
However one can expect that QCD radiative corrections will not change our results 
for the moments  $n=2$ and $n=4$ dramatically. As to the sixth moment, the contribution 
the QCD radiative corrections in this case may be important. 

Based on the values of the moments obtained within QCD sum rules we have proposed the model of LCWF. 
This model has some interesting properties:

1. Due to the evolution (\ref{conf_exp}) the accuracy of the moments obtained within 
the model (\ref{model}) improves as the scale rises. For instance, if the error  
in determination of the moment $\langle \xi^2 \rangle$ is $10 \%$ at scale $\mu=\mu_0=1.2$ GeV, at scale $\mu = 10$ GeV
the error is $4 \%$. For higher moments the improvement of the accuracy is  even better and 
there is no error at all at infinite scale $\mu = \infty$. The improvement of the 
accuracy allows us to expect that the model (\ref{model}) will be rather good even after inclusion 
of the radiative corrections. 

2. At scale $\mu \sim \mu_0$ the LCWF can be considered as wave function of nonrelativistic object 
with characteristic width $\sim v^2 \sim 0.3$. Due to the evolution, at larger scales relativistic tail appears. This tail cannot be considered in the framework of NRQCD and at these scales, strictly speaking,
$\eta_c$ meson  is not a nonrelativistic object. 

3. We have found that due to the presence of high momentum tail in the LCWF at scales $\mu> \mu_0$ 
there is violation of velocity scaling rules obtained in paper \cite{Bodwin:1994jh}.
More exactly, for any scale $\mu > \mu_0$ there is a number $n_0$ for which the moments $\langle \xi^n \rangle, n>n_0$
violate NRQCD velocity scaling rules.

Actually the last two properties are properties of real LCWF of 
$\eta_c$ meson. 

In the last section of our paper we have compared the model of LCWF (\ref{model})
with other model proposed in literature. Thus we have considered the following models:

1. {\bf Asymptotic LCWF (\ref{as}) }. Asymptotic LCWF is wider than LCWF (\ref{model}).
From this comparison we have concluded 
that although $\eta_c$ meson at scale $\mu \sim \mu_0$ can be considered as a nonrelativistic object 
but the width of this meson is rather large and this approximation is not very good. At larger 
scales LCWF of $\eta_c$ is wide and $\eta_c$ cannot be considered as a nonrelativistic object.

2. {\bf The LCWF proposed by Bondar and Chernyak in paper \cite{Bondar:2004sv}. }
The model proposed in paper \cite{Bondar:2004sv} is in good agreement with 
our model of wave function taken at scale $\mu = 10$ GeV.

3. {\bf Effective LCWF proposed by Bodwin, Kang and Lee in paper \cite{Bodwin:2006dm}.}
Effective LCWF of $\eta_c$ meson proposed in paper \cite{Bodwin:2006dm} is considerably 
narrower than our model of LCWF. Moreover in calculation of the cross section 
of the process $e^+ e^- \to J/\Psi + \eta_c$ the authors of paper \cite{Bodwin:2006dm}
did not regard the evolution of their wave function from scale $\mu \sim \mu_0$ 
to characteristic scale of this process $\sim \sqrt s = 10.6$ GeV. We have proposed 
simple model (\ref{mod}) to estimate this effect.  Within this model 
we have shown that the effect is important and good agreement with 
experimental results can be achieved. 

4. {\bf LCWF proposed by Ma and Si (MS) in paper \cite{Ma:2006hc}. }
In this model LCWF is built as a series in $\alpha_s$. At first approximation 
the author took $\sim \delta(\xi)$. Due to the singular next-to-leading order function and infinitely narrow leading
order function LCWF proposed in paper  \cite{Ma:2006hc} has negative moments.

\begin{acknowledgments}
The authors thank G.T. Bodwin for useful  comments on this paper. 
This work was partially supported by Russian Foundation of Basic Research under grant 04-02-17530, Russian Education
Ministry grant RNP-2.2.2.3.6646, CRDF grant MO-011-0, Scientific School grant SS-1303.2003.2. One of
the authors (V.B.) was also supported by the Dynasty foundation and president grant MK-1863.2005.02.

\end{acknowledgments}

\end{document}